\documentclass[11pt]{article}
\usepackage{latexsym}
\usepackage{graphicx, epstopdf}
\usepackage{amsfonts}
\usepackage{amsmath, amsthm, amssymb, mathabx}
\usepackage{fullpage, rotating}
\usepackage{setspace}
\usepackage{longtable, multirow}
\usepackage{natbib}
\bibpunct{(}{)}{;}{a}{}{,}
\usepackage{dcolumn}
\newcolumntype{.}{D{.}{.}{-1}}
\newcolumntype{d}[1]{D{.}{.}{#1}}
\usepackage[top=.8in, left=.5in, right=.5in,
bottom=.8in]{geometry}
\usepackage{bm}
\usepackage[compact]{titlesec}
\usepackage{booktabs}
\usepackage{enumitem}

\usepackage{color}
\RequirePackage[colorlinks,citecolor=blue,urlcolor=blue]{hyperref}
\usepackage{amsthm}

\newtheorem{theorem}{Theorem}

\usepackage{caption}
\usepackage{subcaption}
\usepackage[linesnumbered, ruled, vlined]{algorithm2e}

\expandafter\def\expandafter\normalsize\expandafter{\normalsize\setlength\abovedisplayskip{0pt}}
\expandafter\def\expandafter\normalsize\expandafter{\normalsize\setlength\belowdisplayskip{0pt}}
\expandafter\def\expandafter\normalsize\expandafter{\normalsize\setlength\abovedisplayshortskip{0pt}}
\expandafter\def\expandafter\normalsize\expandafter{\normalsize\setlength\abovedisplayshortskip{0pt}}

\newcommand*\samethanks[1][\value{footnote}]{\footnotemark[#1]}

\begin{document}
\pagestyle{plain}

\title{Estimating optimal tailored active surveillance strategy under interval censoring}

\author{
  Muxuan Liang \thanks{Department of Biostatistics, University of Florida}\\
  \and
  Yingqi Zhao \thanks{Public Health Sciences Division, Fred Hutchinson Cancer Center}\\
  \and
  Daniel W. Lin \thanks{Department of Urology, University of Washington}\\
  \and
  Matthew Cooperberg \thanks{Epidemiology \& Biostatistics, University of California, San Francisco}\\
  \and 
  Yingye Zheng \samethanks[2]\\
}
\date{}
\maketitle
\thispagestyle{empty}

\abstract{
Active surveillance (AS) using repeated biopsies to monitor disease progression has been a popular alternative to immediate surgical intervention in cancer care. However, a biopsy procedure is invasive and sometimes leads to severe side effects of infection and bleeding. To reduce the burden of repeated surveillance biopsies, biomarker-assistant decision rules are sought to replace the fix-for-all regimen with tailored biopsy intensity for individual patients. Constructing or evaluating such decision rules is challenging. The key AS outcome is often ascertained subject to interval censoring. Furthermore, patients will discontinue their participation in the AS study once they receive a positive surveillance biopsy. Thus, patient dropout is affected by the outcomes of these biopsies. In this work, we propose a nonparametric kernel-based method to estimate the true positive rates (TPRs) and true negative rates (TNRs) of a tailored AS strategy, accounting for interval censoring and immediate dropouts. Based on these estimates, we develop a weighted classification framework to estimate the optimal tailored AS strategy and further incorporate the cost-benefit ratio for cost-effectiveness in medical decision-making. Theoretically, we provide a uniform generalization error bound of the derived AS strategy accommodating all possible trade-offs between TPRs and TNRs. Simulation and application to a prostate cancer surveillance study show the superiority of the proposed method.
}

\newcommand{\n}{\noindent}

\newpage

\section{Introduction}
\label{sec:intro}

Active surveillance (AS) has become a popular alternative to immediate aggressive intervention such as surgery in the management of patients diagnosed with low-grade cancer\citep{ganz2012national,cooperberg2015trends,auffenberg2017practice,chen2016active,sanda2018clinically}. It typically involves monitoring tumor progression with invasive testing tools such as biopsy performed periodically with a one-size-for-all schedule for all patients. To reduce the burden of frequent testing, biomarker-assistant rules are sought to provide dynamically tailored AS intervals based on patients' characteristics. Development of these decision rules and assessment of their clinical validities are challenging due to the dynamic nature of AS and how the key AS outcome is ascertained. 

Our research is motivated by the Canary Prostate Active Surveillance Study (PASS), a multicenter, prospective cohort study enrolling men diagnosed with low-grade prostate cancer opting for AS \citep{cooperberg2020tailoring}. In PASS, by the study protocol, patients are monitored closely for disease progression, with repeated prostate-specific antigen (PSA) tests every three months, clinical visits every six months, and ultrasound-guided biopsies at 6, 12, and 24 months after diagnosis, then every two years. A critical goal of this cohort study is to develop an optimally tailored AS dynamic regimen. The outcome of AS is the reclassification from a low-grade disease to a clinically significant disease. The reclassification is probed through a sequence of biopsies; its timing is known only between the last negative biopsy and the most recent biopsy indicating disease progression. The patient will typically drop out of the study once a reclassification is detected. Deriving and evaluating the rule for AS would need to account for the complication of such an outcome subject to interval censoring and immediate dropouts.  

Many model-based approaches have been proposed to characterize the covariate effects on single or multiple interval-censored events. A number of parametric and semiparametric maximum likelihood estimators (MLE) and sieve likelihood estimators were proposed to deal with interval censoring under proportional hazards models \citep{ huang1995maximum, huang1996efficient, rossini1996semiparametric, huang1997sieve, goggins2000proportional, wang2011semiparametric, zeng2017maximum, gao2019semiparametric}, as well as other failure time models such as the additive hazard models and the accelerated failure time (AFT) models \citep{lin1998additive, shiboski1998generalized, shen2000linear, martinussen2002efficient, tian2006accelerated, lin2010semiparametric}. To construct surveillance rules with longitudinal measurements, approaches such as joint modeling or partly conditional models can be adapted with these baseline models to account for interval-censored outcomes \citep{tsiatis2004joint, yu2008individual, tomer2019personalized,maziarz2017longitudinal}. These approaches typically involve first estimating the dynamic risks under parametric/semiparametric model assumptions and then constructing rules with the risk threshold to achieve a specific objective. However, these approaches rely on the specified model assumptions, and their performance may be susceptible to the required assumptions. In addition, these approaches often require substantial computational efforts (e.g., expectation–maximization algorithms) \citep{mongoue2008new, mcmahan2013regression}. Thus, a robust treatment for the interval-censored event under a more flexible and computationally efficient framework would broaden the applicability of the developed rules.

Recently alternative classification-based approaches have been adopted in deriving decision rules in medical decision-making. \cite{dong2023constructing} proposed a classification framework incorporating both the time-dependent longitudinal true positive rate (TPR) and the true negative rate (TNR) in the objective function for learning the optimally tailored dynamic surveillance rules. Their work accommodated scenarios with right censored outcomes, using an inverse-censoring-probability weighted (IPCW) method to estimate the combined TPR and TNR objective function. Then a computationally efficient algorithm was developed to estimate the optimal rules, leveraging existing machine learning algorithms and convex optimization \citep{bartlett2006}. However, for settings where outcomes are detected less frequently through diagnostic procedures, the approach, especially the IPCW estimation method, does not directly apply due to interval censoring. \cite{chan2021developing} proposed non-parametric estimators for TPR and TNR via kernel regressions to evaluate the prediction performance of a baseline risk score when the occurrence of a particular clinical condition is only examined at several prescheduled visit times. Their proposed estimators are valid regardless of the adequacy of the fitted model given rise to the risk score, and computationally easy to implement. However, they assume completely random dropouts, which may not be true for a surveillance study where patients may drop out from the study immediately after detecting disease progression based on the results of procedures.

In this work, in the context of deriving clinical decision rules directly optimizing classification-based objective function, we develop a flexible framework that can handle interval-censored events and non-random dropouts with computationally efficient algorithms. We make two major contributions. Firstly, to evaluate any given tailored AS strategy, we propose nonparametric estimators for TPR and TNR using nonparametric kernel estimators, where a two-dimensional kernel function is adopted to deal with both interval-censored events and non-random dropouts simultaneously. Secondly, to construct the optimal AS strategy, we maximize a weighted benefit value function, accounted for such complications in outcome ascertainment. The value function involves weighted benefits using the cost-benefit ratios and nonparametrically estimated prevalence as weights. To accommodate these weights involving estimates, in our theoretical study, we provide a uniform bound for the generalization errors of the derived AS decision rules under all possible choices of the weights associated with the value function. Simulations and an application to PASS are conducted to show the superiority of the proposed method.

The remainder of the paper is organized as follows. In Section~\ref{sec:modeling}, we introduce our general decision-making framework and the proposed estimation method. In Section~\ref{sec:theory}, we provide the theoretical results. In Section~\ref{sec:sim}, we present the simulation results demonstrating the finite sample performance of the proposed method. In Section~\ref{sec:real}, we apply our proposed method to the PASS data and evaluate the derived decision rule on an external data source. To conclude, in Section~\ref{sec:discuss}, we discuss possible methodological extensions to this general decision-making framework for cancer surveillance.
	
\section{Method}
\label{sec:modeling}

\subsection{Weighted benefits value function and the optimality}
\label{subsec:value}

Let $Z_t$ represent the covariate information at time $t$, $\{Z_t\}_{t\in \mathbb{R}_+}$ be a $p$-dimensional covariate process, and $\widebar{Z}_t$ represent the accrued covariate information up to $t$. Our goal is to derive a tailored AS decision rule, $d_s(\cdot)$, which maps $\widebar{Z}_s$, the accrued information up to the decision time point $s$, to a binary output $\{1,-1\}$,  with $d_s(\cdot) = 1$ indicating a positive decision for conducting a future surveillance biopsy at $s+\tau$, and $d_s(\cdot) = -1$ for a decision to skip the biopsy at that time. Here $\tau$ is typically predetermined by the study protocol fixed for everyone. Therefore, $d_s(\cdot)$ will lead to a tailored surveillance intensity dependent on the individual's covariate history. In particular, for ease of implementation and stable estimation given a typical limited study cohort size, we are interested in the stabilized strategy $d_0(\cdot)$, i.e., $d_s(\widebar{Z}_s)=d_0(\widebar{Z}_s)$. A stabilized strategy shares the same format at different time points $s$, and takes only the most up-to-date covariate information as input. 

The validity of $d_s(\cdot)$, i.e., whether a biopsy should be scheduled at time $s+\tau$, depends on whether a surveillance endpoint will occur within the time window $[s, s+\tau]$. For any tailored AS rule, we first define a weighted benefits value function, which is based on the TPR and the TNR \citep{dong2023constructing}. At a landmark time point $s$, pertinent to the outcome by a future time $s+\tau$, the time-varying TPR and TNR for a tailored AS strategy $d_s(\widebar{Z}_s)$ are defined as
$
	\text{TPR}(d_s;s,\tau)=P\left\{d_s(\widebar{Z}_s)=1\mid s<T\leq s+\tau\right\}
$ and $
	\text{TNR}(d_s;s,\tau)=P\left\{d_s(\widebar{Z}_s)=-1\mid T> s+\tau\right\}.
$
The $\text{TPR}(d_s;s,\tau)$ is the proportion of positive decisions among patients with an AS event occurs within time interval $(s,s+\tau]$; the $\text{TNR}(d_s;s,\tau)$ is the proportion of negative decisions among patients who are event-free by $s+\tau$. Both high $\text{TPR}(d_s;s,\tau)$ and $\text{TNR}(d_s;s,\tau)$ are desirable for meaningful clinical decisions, but there is often a tradeoff between the two. We therefore define the time-specific weighted benefits value function at time point $s$ as
$
	\phi(d_s;s,\xi,\tau)=\text{TPR}(d;s,\tau) +\xi(s) \text{TNR}(d;s,\tau),
$
where $\xi(s)$ is a pre-specified scalar representing the trade-off between $\text{TPR}(d;s,\tau)$ and $\text{TNR}(d;s,\tau)$. To obtain a dynamic regimen over time, we define the weighted benefits value function by averaging time-specific value functions over all landmark time points. Let $\{S(t)\}_{t\in \mathbb{R}_+}$ represent the sampling process of the biopsy decision landmark time points, i.e. $S(t)$ is the distribution function of the time making biopsy decisions. The value function is defined as
$
	\Phi\left(\bm d;\xi,\tau\right):=\int \phi(d_t;t,\xi(t))\mathrm{d} S(t),
$
where $\bm d=\left\{d_s\right\}_{s\geq 0}$.

Based on the definition of the weighted-benefit value function, the optimally tailored AS regimen under a specific $\xi(\cdot)$ is then defined as the maximizer of the value function, i.e.
$
	\bm d_{\xi,\tau}:=\arg\max \Phi(\bm d;\xi,\tau).
$
When the biopsy decisions have to be made at fixed landmark decision time points denoted as $0\leq t_1 <t_2 <\cdots<t_J$, the value function
$
	\Phi(\bm d;\xi,\tau)=J^{-1}\sum_{j=1}^J \phi(d_{t_j};t_j,\xi(t_j),\tau).
$
If we are interested in stabilized decision rule, the weighted-benefit value function can be written as
$
	\Phi(d_0;\xi,\tau)=J^{-1}\sum_{j=1}^J \phi(d_0;t_j,\xi(t_j),\tau).
$

There are many possible choices of $\xi(\cdot)$. One of the possible choices is to specify a $\xi(s)$ that characterizes the cost-benefit trade-offs. In this case, a strategy is cost-effective at time $s$ if the number of unnecessary biopsies a patient can afford to catch an event (disease progression) is lower than an expected number, referred to as $r$ \citep{pepe2016early}. It can be achieved by choosing $\xi(s)=(1-\rho(s;\tau)/(\rho(s;\tau) r)$, where $\rho(s;\tau)=P\left\{s<T\leq s+\tau\mid T>s\right\}$.


\subsection{Estimating optimally tailored regimen under interval censoring and immediate dropouts}
\label{subsec:est_tnr_tpr}

In this section, we consider the estimation of the time-varying TPR/TNR and the optimal tailored AS strategy using the observed data. To begin with, we first introduce our notations and assumptions for the observed data.

Denote the event and censoring times as $T$ and $C$, respectively. In the observed data, we do not directly observe $T$; instead, physicians would set up $K$ biopsies at prespecified times $N=(N_1,\cdots, N_K)$,  where $N_1< \cdots< N_K$, to check whether there is disease progression. Given these biopsy time points, without missing data or dropouts, we observe $\Delta=(\Delta_1,\cdots, \Delta_K)$, where $\Delta_k=1\{T\leq N_k\}$ indicating whether the disease progressed before the $k$-th biopsy. However, we may be unable to observe $\Delta_k$ and $N_k$ due to lost-to-followup before the event time (censoring), missed biopsy appointments, and dropout due to disease progression. Specifically, to account for possible missed biopsy appointments, we use $\delta=(\delta_1, \cdots, \delta_K)$ to indicate the completeness of the biopsy sequence, where $\delta_k=1$ indicating information on the $k$-th biopsy, as well as $\Delta_k$, is available. To account for the censoring before the event time, let $\zeta=(\zeta_1\cdots, \zeta_K)$, where $\zeta_k=1\{C>N_k\}$ indicating whether the censoring time is later than the $k$-th biopsy time, i.e., the $k$-th biopsy is not censored; if $\zeta_k=0$, we cannot observe the $k$-th biopsy, $N_k$ and $\Delta_k$ either. In addition, we assume that the patient will drop out of the study immediately after $\Delta_k=1$. Under these notations, in our observed data, we can observe $N_k$ and $\Delta_k$ if and only if $\zeta_k\delta_k=1$ and $\Delta_{k'}\delta_{k'}=0$ for all $k'<k$.

For $N$, $\zeta$, and $\delta$, we adopt the same assumptions as those in \cite{chan2021developing}. We assume that $N$ is a random vector as patients may visit at random times near the scheduled visits, i.e., the biopsy times $N$ are independent of both $T$ and $\{Z_t\}_{t\in \mathbb{R}_+}$; the $P(\delta_k=1\mid \Delta, N, \{Z_t\}_{t\in \mathbb{R}_+})=\rho_k>0$; the censoring indicator $P(\zeta_k=1\mid \Delta, N, \{Z_t\}_{t\in \mathbb{R}_+})=\widetilde{\rho}_k>0$. The key difference between the settings in \cite{chan2021developing} and ours is whether the patient will drop out from the study immediately after $\Delta_k=1$. For settings in \cite{chan2021developing}, it is possible that the patients may still return to the study after $\Delta_k=1$; for surveillance study, the patients often drop out from the study and seek other medical interventions once $\Delta_k=1$ for some $k$.

Next, we propose an estimation method of the time-varying $\text{TNR}(d_s;s,\tau)$ based on the observed data under a tailored AS strategy, $d_s$. Following the approach in \cite{chan2021developing}, we can construct a nonparametric estimation for time-varying $\text{TNR}(d_s;s,\tau)$ for a given decision rule $d_s$. The key idea is to leverage the randomness of the biopsy time. Given an event time, since the biopsy time is random, there are chances that the event time is right before one biopsy of one individual and right after another biopsy of another similar individual. Thus, by combining the biopsy information across similar individuals, we can estimate the TPR/TNR. 

Define
$
	F_a(t;s)=P\left\{d_s(\widebar{Z}_s)=a, T>t\right\},
$
where $a=\{1,-1\}$. The $\text{TNR}(d_s;s)$ can be re-formulated as a function of $F_a(t;s)$, i.e.,
$
	\text{TNR}(d_s;s)=F_{-1}(s+\tau;s)\{F_{-1}(s+\tau;s)+F_1(s+\tau;s)\}^{-1}.
$
Following \cite{chan2021developing}, we consider the following estimation for $\text{TNR}(d_s;s)$, i.e.,
$
	\widehat{\text{TNR}}(d;s)=E_n[1\{d(\widebar{Z}_s)=-1\}W_{-1,s+\tau}],
$
where
$
	W_{-1,t}=\{\sum_k(1-\Delta_k)\zeta_k\delta_kK_h(N_k-t)\}\{\sum_kE_n[(1-\Delta_k)\zeta_k\delta_kK_h(N_k-t)]\}^{-1},
$
the function $K_h(\cdot)=h^{-1}K(\cdot/h)$ and $K(\cdot)$ is a univariate kernel function, and $h$ is the bandwidth. The proposed estimator utilizes all observed negative biopsies. Although we do not observe future positive biopsy results after a positive biopsy, we observe all negative biopsies except those missing or censored. Thus, the proposed estimator for time-varying $\text{TNR}(d_s;s,\tau)$ is also expected to be consistent in our setting.


However, the estimation of $\text{TPR}(d_s;s,\tau)$ is not straightforward. In our setting, patients immediately drop out from the study once $\Delta_k=1$ for some $k$, and thus the positive biopsy times after the first positive biopsy cannot be observed. If directly using the estimator in \cite{chan2021developing} for TPRs in our settings, it leads to a biased estimation since whether we can observe a positive biopsy also depends on previous biopsy results. To address the immediate dropouts, we consider adjacent negative-positive pairs of biopsies. We say an adjacent pair of biopsies as a negative-positive pair if and only if $\Delta_{(k)}=0$ and $\Delta_k=1$, where $(k)$ is the index of the adjacent observed biopsy before the $k$-th biopsy. Different from the positive biopsies, whether an adjacent negative-positive pair will be observed does not depend on the past biopsy results; the adjacent negative-positive pair will always be observed if there is no censoring or missing, and thus is not affected by the immediate dropouts. Thus, the biopsy times of an adjacent negative-positive pair can always inform the shortest interval identifiable from the observed data that contains the event time, i.e., the event happens within $(N_{(k)}, N_k]$. In addition, since the biopsy times are random, the biopsy times of adjacent negative-positive pairs are random. Given an event time, there are chances that the event time is right before the positive biopsy time of one pair, and right after the negative biopsy time of another pair. Thus, by combining information of multiple adjacent negative-positive pairs across similar individuals, we can address the problem of interval censoring. Denote $N_0=0$ and $\zeta_0\delta_0=1$, which corresponds to the confirmatory biopsy or baseline diagnosis. Theorem~\ref{thm:adjacent_biopsy} shows that the $P\left\{d_s(\widebar{Z}_s)=a, s \leq T\leq s+\tau\right\}$ is identifiable using observed adjacent negative-positive pairs. Its proof can be found in the online Supporting Information.

\begin{theorem}\label{thm:adjacent_biopsy}
	For any $k$ and $s$, we have
	\begin{eqnarray*}
		&&P\left\{d_s(\widebar{Z}_s)=a, s \leq T\leq s+\tau\right\}\\
		&=&P\left\{d_s(\widebar{Z}_s)=a, \Delta_{(k)}=0, \Delta_k=1\mid N_{(k)}=s, N_k=s+\tau, \delta_k\zeta_k=1\right\},
	\end{eqnarray*}
	where $(k)$ is the index of the adjacent observed biopsy before the $k$-th biopsy.
\end{theorem}


Following Theorem~\ref{thm:adjacent_biopsy}, for any $k$, notice that
\begin{eqnarray*}
    \text{TPR}(d_s;s,\tau)&=&\frac{P\left\{d_s(\widebar{Z}_s)=a, \Delta_{(k)}=0, \Delta_k=1, \delta_k\zeta_k=1\mid N_{(k)}=s, N_k=s+\tau\right\}}{P\left\{\Delta_{(k)}=0, \Delta_k=1, \delta_k\zeta_k=1\mid N_{(k)}=s, N_k=s+\tau\right\}}.
\end{eqnarray*}
Thus, the $\text{TPR}(d_s;s,\tau)$ can be then estimated by
$
	\widehat{\text{TPR}}(d_s;s,\tau)
	=E_n[1\{d_s(\widebar{Z}_s)=1\}W_{1,s}],
$
where 
$
	W_{1,t}=\{\sum_k \Delta_k(1-\Delta_{(k)})\zeta_k\delta_k\widetilde{K}_{\widetilde{h}}(N_k-s-\tau, N_{(k)}-s)\}\{\sum_k E_n[\Delta_k(1-\Delta_{(k)})\zeta_k\delta_k\widetilde{K}_{\widetilde{h}}(N_k-s-\tau, N_{(k)}-s)]\}^{-1},
$
the function $\widetilde{K}_{\widetilde{h}}(t_1,t_2)=\widetilde{h}^{-2}\widetilde{K}(t_1/\widetilde{h},t_2/\widetilde{h})$, $\widetilde{K}(\cdot,\cdot)$ is a two-dimensional kernel function, and $\widetilde{h}$ is the associated bandwidth that could be different from $h$. 

Based on the estimators of $\text{TNR}(d_s;s,\tau)$ and $\text{TPR}(d_s;s,\tau)$, we can estimate the optimal tailored AS strategy. For the simplicity of the notation, we only consider the strategy with a stabilized decision rule in the following discussion. For stabilized decision rules, we can maximize
$
	\widehat{\Phi}_n(d_0;\xi,\tau)=J^{-1}\sum_j E_n[1\{d_{0}(\widebar{Z}_{t_j})=1\}W_{1,t_j}+1\{d_{0}(\widebar{Z}_{t_j})=-1\}\xi(t_j)W_{-1,t_j+\tau}].
$

\subsection{Computationally efficient algorithms}
\label{subsec:est_rule}

Maximizing the weighed benefits value function is equivalent to solving a weighted classification problem, i.e.,
$
    \min_{\bm d} J^{-1}\sum_j E_n[1\{d_{0}(\widebar{Z}_{t_j})\not=1\}W_{1,t_j}+1\{d_{0}(\widebar{Z}_{t_j})\not=-1\}\xi(t_j)W_{-1,t_j+\tau}].
$
 
 To prevent the complication of optimizing an objective function that includes the indicator function, we substitute it with a convex surrogate loss function, denoted as $\phi$, and consider
\begin{eqnarray}\label{eq:opt_sample}
	\min_{f\in \mathcal{F}} \ell_{\phi,n}(f;\xi, \lambda_n)=J^{-1}\sum_j E_n\left[W_{+,t_j}\phi\left\{f(\widebar{Z}_{t_j})\right\}+W_{-,t_j}\phi\left\{-f(\widebar{Z}_{t_j})\right\}\right]+\lambda_n\|f\|_{\mathcal{F}}^2,
\end{eqnarray}
where $W_{+,t_j}=\left[W_{1,t_j}-\xi(t_j)W_{-1,t_j+\tau}\right]_{+}$ and $W_{-,t_j}=\left[W_{1,t_j}-\xi(t_j)W_{-1,t_j+\tau}\right]_{-}$, $\mathcal{F}$ is a pre-specified function class in a Hilbert space, and $\|\cdot\|_{_{\mathcal{F}}}$ is the associated norm. The penalization $\lambda_n\|f\|_{\mathcal{F}}^2$ is added to avoid over-fitting, where $\lambda_n$ is a tuning parameter. Denote its minimizer as $\widehat{f}_{\xi,\lambda_n}$; the estimated AS strategy can be characterized by the stabilized decision rule $\widehat{d}_{\xi,\lambda_n}(\widebar{Z}_t)=\mathrm{sgn}\left\{\widehat{f}_{\xi,\lambda_n}(\widebar{Z}_t)\right\}$.

In our formulation, to account for cost-benefit ratios, we choose $\xi(s)=(1-\rho(s;\tau)/(\rho(s;\tau) r)$. For constructing the objective function, it's necessary to estimate $\xi(s)$. In the online Supporting Information, we derive an estimator for $\xi(s)$ by applying similar techniques used in constructing $\widehat{\text{TPR}}(d_0;s,\tau)$. Denote the estimated $\xi(s)$ as $\widehat{\xi}(s)$, and we can minimize $\ell_{\phi,n}(f;\widehat{\xi}, \lambda_n)$ over $f\in \mathcal{F}$. Denote its minimizer as $\widehat{f}_{\widehat{\xi}, \lambda_n}$; the estimated AS strategy is defined by $\widehat{d}_{\widehat{\xi},\lambda_n}(\widebar{Z}_t)=\mathrm{sgn}\left\{\widehat{f}_{\widehat{\xi},\lambda_n}(\widebar{Z}_t)\right\}$.

Minimizing $\ell_{\phi,n}(f;\xi, \lambda_n)$ fundamentally resolves a weighted classification problem through the use of penalized empirical risk minimization. As the objective function $\ell_{\phi,n}(f;\xi, \lambda_n)$ is convex in $f$, we can employ the gradient-based approaches for its solution. If $\phi$ is chosen as the logistic loss with linear decision rules (i.e., $f(\cdot)$ has a linear form), minimizing $\ell_{\phi,n}(f;\xi, \lambda_n)$ is the same as a weighted logistic regression with a ridge penalty. Existing R packages, e.g., \textit{glmnet}, can be used to implement the proposed method. We refer to our proposed method as the Optimization with the Surrogate Function approach for Interval-censored data (OSF-I).
	
\section{Theoretical properties}
\label{sec:theory}

In this section, we state the theoretical properties of the proposed estimators under a stabilized decision rule. The detailed proof of the main theorem can be found in the online Supporting Information. The theoretical properties of the time-varying surveillance decision rules are implied in the proof. To start with, given an decision rule $d_0$, we define 
$
{\Phi}(d_0;\xi,\tau)=J^{-1}\sum_j \left\{\text{TPR}(d_0;t_j,\tau)+\xi(t_j)\text{TNR}(d_0;t_j,\tau)\right\},
$
where $\xi=(\xi(t_1), \cdots, \xi(t_J))$. To assess the theoretical property of the tailored AS rule under $\widehat{d}_{\xi, \lambda_n}$, we use a generalization error that compares ${\Phi}(\widehat{d}_{\xi, \lambda_n};\xi,\tau)$ with the optimally tailored AS dynamic regimen. The optimally tailored AS dynamic regimen at time $t_j$ is defined as the maximizer of $\text{TPR}(d;t_j,\tau)+\xi(t_j)\text{TNR}(d;t_j,\tau)$. Denote the maximizer at time $t_j$ as $d^*_{\xi, j}$, and define $\Phi^*(\xi,\tau)=J^{-1}\sum_j \left\{\text{TPR}(d^*_{\xi, j};t_j,\tau)+\xi(t_j)\text{TNR}(d^*_{\xi, j};t_j,\tau)\right\}.$ The generalization error is then defined as
$
	\Phi(\widehat{d}_{\xi, \lambda_n};\xi,\tau)-{\Phi}^*(\xi,\tau).
$
To accommodate the case where $\xi$ is chosen using the cost-benefit ratio, we derive an upper bound for the generalization error $\left\{\Phi(\widehat{d}_{\xi, \lambda_n};\xi,\tau)-{\Phi}^*(\xi,\tau)\right\}$ which is uniformly held for $\xi\in \Xi:=[\underline{\xi},\widebar{\xi}]^J$, where $\underline{\xi}$ is some constant bounded away from $0$ and $\widebar{\xi}$ is some constant bounded away from $+\infty$. 

For the function class $\mathcal{F}$, we impose a complexity constraint regarding the covering number of the space $\mathcal{F}$. The covering number $N\{\epsilon, \mathcal{F}, L_2(P)\}$ is defined as the minimal number of closed $L_2(P)$-balls of radius $\epsilon>0$ required to cover $\mathcal{F}$, where $\|f\|_{P,2}^2=E[f^2]$ \citep{vandegeer2008}. Under these notations, we assume the following assumptions:
\begin{enumerate}
\item\label{cond:complexity} There exists constants $0<v<2$ and $c$ such that $\forall \epsilon\in (0,1]$, we have $\sup_P\log N\left\{\epsilon, \mathcal{F}, L_2(P)\right\}\leq c\epsilon^{-v}$, where the supremum is taken over all finitely discrete probability measures $P$.
\item\label{cond:kernel} The kernel function $K(\cdot)$ is a $\nu$th order uni-variate kernel function with a bounded 2nd order derivative and compact support; the kernel function $\widetilde{K}(\cdot, \cdot)$ is a $\nu$th order bivariate kernel function with a bounded 2nd order derivative and compact support.
\end{enumerate}
Assumption~\ref{cond:complexity} controls the complexity of the function class $\mathcal{F}$ and can be satisfied for many choices of function classes. For example, if $\mathcal{F}$ is a class of all linear combinations of elements in a fixed base class with a finite Vapnik-Chervonenkis dimension, Assumption~\ref{cond:complexity} is satisfied according to Theorem~9.4 in \cite{kosorok2008introduction}. Assumption~\ref{cond:kernel} contains commonly adopted assumptions for kernel regressions \citep{nadaraya1964estimating}. Assumptions on the surrogate loss $\phi$ can be found in the online Supporting Information.

Under these assumptions, our main theorem below provides a uniformly valid upper bound for the generalization error.
\begin{theorem}\label{thm:generalization_error}
	Suppose that Assumptions~\ref{cond:complexity} and~\ref{cond:kernel} hold with $\lambda_n\to 0$, with probability approaching to $1$, we have that
	$
		\Phi(\widehat{d}_{\xi, \lambda_n};\xi)-\Phi^*(\xi)\lesssim J^{-1/s}C\left[\mathcal{A}(\lambda_n;\xi)+\lambda_n^{-1/2}\hbar\right]^{1/s}
	$
	uniformly holds for all $\xi\in \Xi$,
	where
	$
		\mathcal{A}(\lambda_n;\xi)$ is the approximation error due to the function class $\mathcal{F}$ (see formula in the online Supporting Information), $
		\hbar=h^{\nu}+(nh)^{-1/2}+\widetilde{h}^{\nu}+(n\widetilde{h}^2)^{-1/2},
	$ and $s$ is a positive constant depending on the choice of $\phi$.
\end{theorem}

The result in Theorem~\ref{thm:generalization_error} shows an upper bound for the weighted benefits value difference between the estimated tailored AS rule and the optimally tailored AS dynamic regimen. 
To achieve the lowest generalization error, we can set $h$ and $\widetilde{h}$ to minimize $\lambda_n^{-1/2}\hbar$; when $h=n^{-1/(2\nu+1)}$ and $\widetilde{h}=n^{-1/(2\nu+2)}$, the term $\lambda_n^{-1/2}\hbar$ is minimized for any given $\lambda_n$. In our simulation and real data, we specify $h=n^{-1/5}$ and $\widetilde{h}=n^{-1/6}$ to reduce the number of tuning parameters. To select the optimal $\lambda_n$, we use the cross-validation procedure. From the uniform generalization error, if we adopt $\widehat{\xi}(\cdot)$ as $\xi(\cdot)$ in optimization~\eqref{eq:opt_sample}, then we can provide a generalization error for $\Phi(\widehat{d}_{\widehat{\xi},\lambda_n};\xi^*)$ (see online Supporting Information), which is the generalization error of the estimated tailored AS rule incorporating the cost-benefit ratio.

\section{Simulations}
\label{sec:sim}

In this section, we compare the proposed method to estimate TPR, TNR, and the tailored AS rule with other methods via simulations. 

\subsection{Data generation}

The data generating process is as follows. We first generate the underlying covariate with measurement error, i.e. $X_l(t)=W_l(t)+\epsilon_l(t)$, where $W_l(t)=a_{0,l}+a_{1,l}\log (t/\nu)$ and $a_l=(a_{0,l},a_{1,l})$ are generated from a bivariate normal distribution with mean $(-0.1,-0.1)^\top$ and covariance matrix $(0.82^2, -0.005;-0.005, 0.13^2)$. The measurement errors $\epsilon_l(t)$ are independently generated from a mean-zero gaussian distribution with a variance of $0.1$. Then we generate the true event time, censoring time, and biopsy information following two scenarios. 
\begin{enumerate}
	\item \label{scenario1} The true event time $T$ follows a proportional hazard model $\lambda(t)=\lambda_0(t)\exp(-0.7W_2(t)+0.8W_3(t)-1.3W_4(u))$, where the baseline hazard $\lambda_0(t)=t/\nu(t/\nu_{\rm scale})^{\nu_{\rm shape}-1}$ and $\nu=30$, $\nu_{\rm scale}=15$, and $\nu_{\rm shape}=1.4$. The censoring time $C$ is generated from a uniform distribution on $[12,150]$.
	\item \label{scenario2} The true event time $T$ is generated from 
$
 	12+\nu\left(\widetilde{T}\nu_{\rm shape}^{-1} \gamma\exp\{-a_{0,\cdot}^\top\beta-r(a_{0,1}+a_{0,2})^2\}\right)^{1/\gamma},
$
	 where $\widetilde{T}$ follows a standard exponential distribution, $\gamma=\nu_{\rm shape}+a_{1,\cdot}^\top\beta+r(a_{1,1}+a_{1,2})^2$, $r=0.1$, $\nu=30$, $\nu_{\rm scale}=15$, and $\nu_{\rm shape}=1.4$. The censoring time $C$ is generated from a uniform distribution on $[12,150]$. 
\end{enumerate}
Scenario~(1) and Scenario~(2) are different in the distributions of $T$. In Scenario~(1), the log-hazard model is linear in $W_l(t)$; in Scenario~(2), the log-hazard model is not linear in $W_l(t)$. By comparing the results in linear and non-linear settings, we examine whether the advantage of the proposed method over existing methods are robust to additional non-linear terms. For both scenarios, we generate the biopsy time depending on the biopsy gap which controls the frequency/intensity of the biopsies. Let $T_{\text{gap}}$ be the biopsy gap. The first biopsy is generated from a uniform distribution on $[12, 3T_{\text{gap}}]$. After the first biopsy time $N_1$, we generate the rest biopsies sequentially. The following biopsy $N_t$ is generated from a uniform distribution on $[N_{t-1}+T_{\text{gap}}, N_{t-1}+3T_{\text{gap}}]$ until $N_{t-1}+T_{\text{gap}}>150$, where $N_{t-1}$ is the previous biopsy time. Through this generation process, the first follow-up biopsy $N_1$ is ensured after 12 months of confirmatory biopsy $N_0=0$; and the adjacent biopsies have a minimum gap of $T_{\text{gap}}$. Then we generate $\Delta=(\Delta_1,\cdots, \Delta_K)$, where $\Delta_k=1\{T\leq N_k\}$; and $\zeta=(\zeta_1\cdots, \zeta_k)$, where $\zeta_k=1\{C>N_k\}$.

\subsection{Comparison between estimators of TPRs and TNRs}
\label{subsec:eval}

In this section, we compare the proposed method for estimating TPRs and TNRs, referred to as `KR-I', and other approaches in terms of the estimation of TPRs and TNRs. We consider the IPCW method used in \cite{dong2023constructing} (referred to as `IPCW') and the method proposed in \cite{chan2021developing} (referred to as `KR-CS'). To implement the IPCW method, we use the Kaplan-Meier estimator to estimate the distribution of the censoring time and use the inverse-censoring-probability weighted method to estimate TPRs and TNRs. 

\begin{figure}
	\centering
	\includegraphics[width=0.8\linewidth]{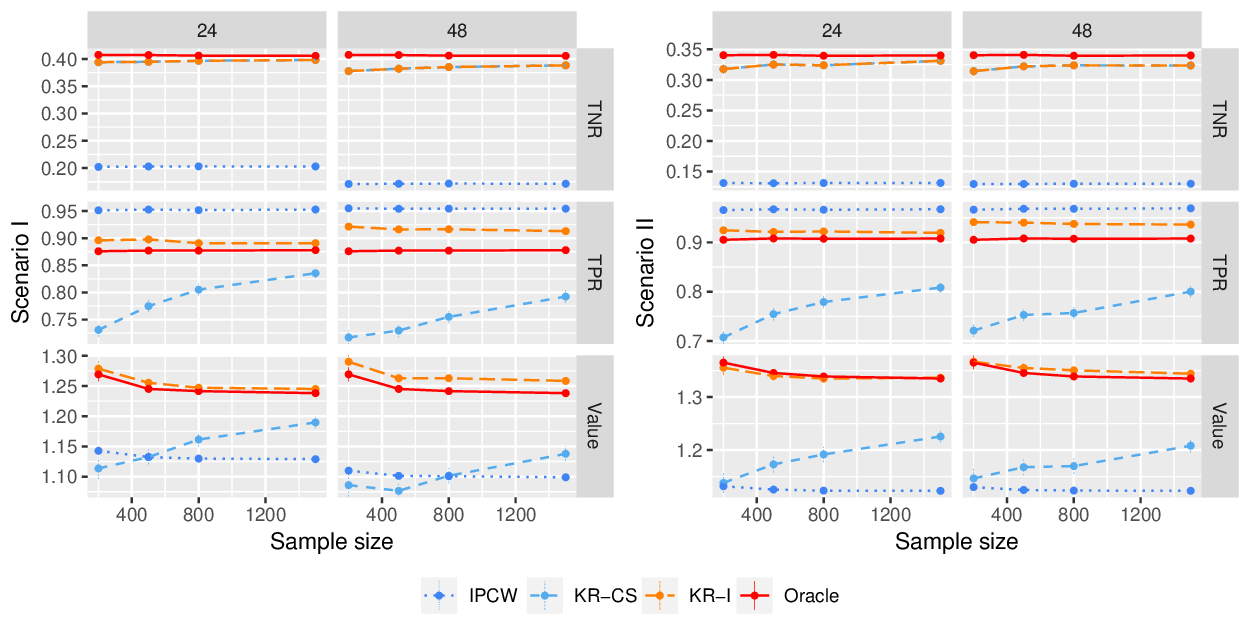}
	\caption{Estimating TNR, TPR, and weighted benefits value of a derived tailored AS rule using IPCW method in \cite{dong2023constructing} (`IPCW'), the method proposed in \cite{chan2021developing} (`KS-CS'), and our proposed method (`KS-I'). The lines labeled as ``Oracle" are the TPRs, TNRs, and values calculated using the true event time without censoring.}
	\label{fig:eval}
\end{figure}

To show the performance of different approaches, we estimate TPRs and TNRs of a fixed surveillance rule using different approaches. To derive the tailored AS rule and generate datasets to estimate TPRs and TNRs, we assume that there is no missed biopsy and the biopsy gap $T_{\text{gap}}$ is $24$. We generate a dataset with a sample size of $500$, and use the optimization with the surrogate function approach for right-censored data (referred to as `OSF-R') proposed in \cite{dong2023constructing} to derive a tailored AS rule (fixing $r=3$ in Scenario~(1); and $r=2$ in Scenario~(2)). To evaluate the derived surveillance rule, we generate an independent dataset with sample sizes varying from $200$ to $500$ and implement our proposed KR-I method, the IPCW method, and the KR-CS approach. When generating this dataset, we vary the biopsy gap $T_{\text{gap}}$ from $24$ to $48$. We use the true event time $T$ to calculate the true TPRs, TNRs, and weighted benefits values of the derived tailored AS rule. The procedure of evaluating the derived surveillance rule using different approaches is repeated $500$ times. Figures~\ref{fig:eval} summarizes the results for Scenarios~(1) and~(2). In both scenarios, the proposed method achieves most accurate estimates w.r.t the true TPRs, TNRs, and weighted benefits values. 

\subsection{Comparison between methods to estimate the tailored AS rule}
\label{subsec:rule}

In this section, we compare the proposed method (referred to as `OSF-I') and the OSF-R approach to estimate the optimal tailored AS rule. The OSF-R approach minimizes a relaxation of the empirical objective that is similar to our objective. However, the OSF-R treats the event time as the biopsy time subjective to right-censoring, and employs an IPCW method to account for the right-censoring. As shown in Section~\ref{subsec:eval}, the IPCW method can lead to biased estimations in TPRs and TNRs, and thus their method may lead to a biased estimation in the optimal tailored AS rule. 

For each scenario, we vary the sample size from $200$ to $500$, and the biopsy gap $T_{\text{gap}}$ from $24$ to $48$; we also vary the $r$ from $2$,$4$,$6$ to $8$. The varies in the sample size, count of biopsy, $\xi$ and scenarios lead to in total $32$ simulation settings. For each simulation setting, we generate the training data and estimate the decision rule on the training data; we repeat this procedure $500$ times. To compare different methods, we then generate an independent testing dataset with a sample size of $1000$. On the testing dataset, we record the true event time $T$; thus, we can directly calculate the TPR, TNR, and the value of the weighted net benefit based on the recorded true event time for each derived AS rule. 

\begin{figure}
	\centering
	\includegraphics[width=0.5\linewidth]{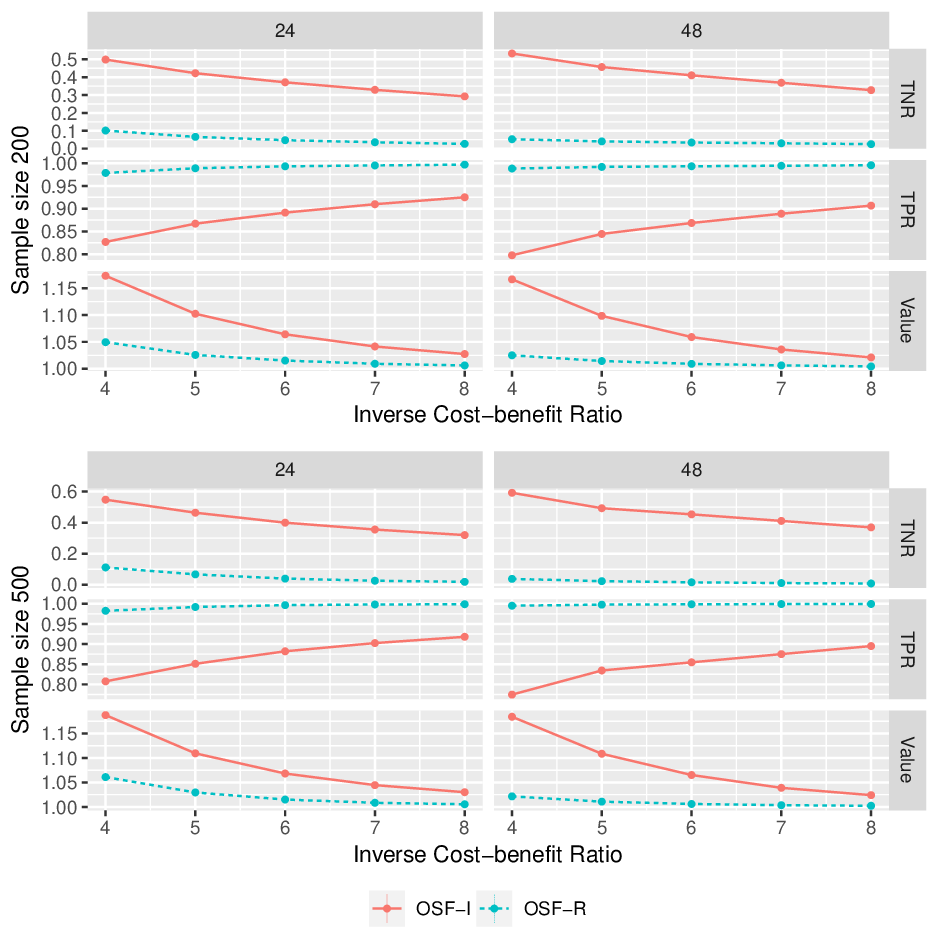}
	\caption{TNR, TPR, and weighted benefits value achieved by different methods for Scenario~(1). The x-axis represents the inverse cost-benefit ratio, $r$, i.e., how many unnecessary biopsies the patient can afford to catch an event (disease progression). The left and right columns summarize the results where $T_{\text{gap}}=24$ and $T_{\text{gap}}=48$, respectively.}
	\label{fig:sim1}
\end{figure}

\begin{figure}
	\centering
	\includegraphics[width=0.5\linewidth]{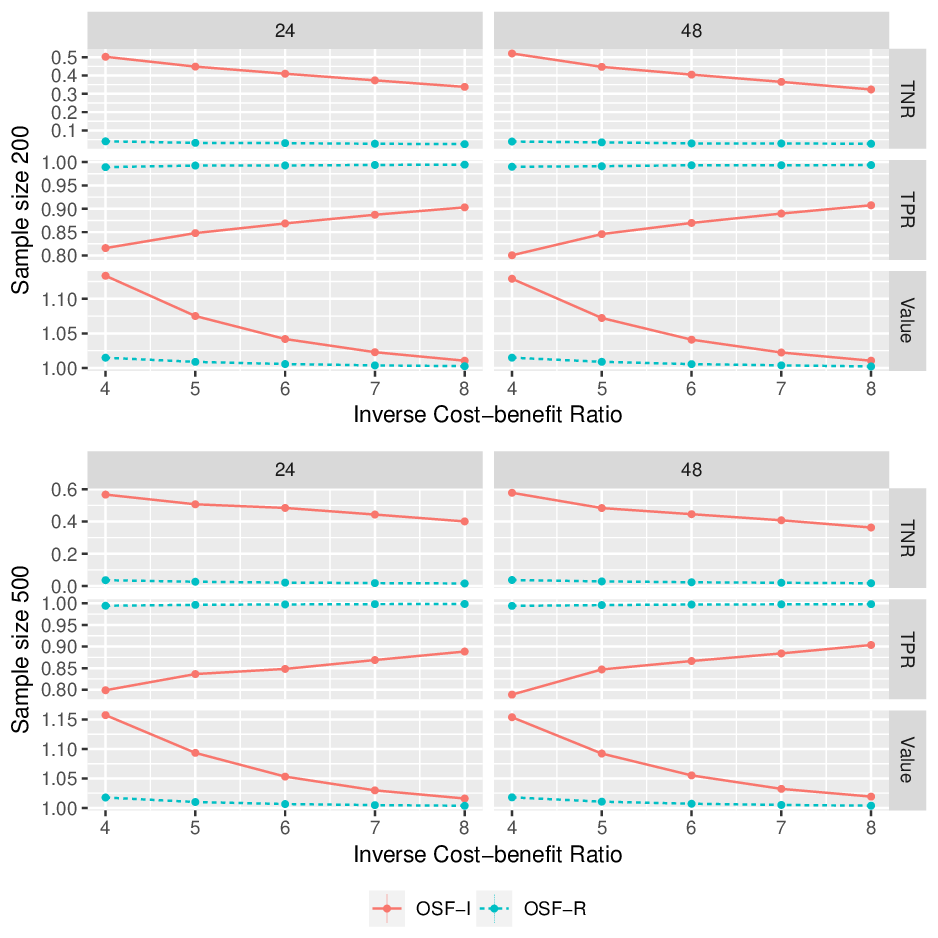}
	\caption{TNR, TPR, and weighted benefits value achieved by different methods for Scenario~(2). The x-axis represents the inverse cost-benefit ratio, $r$, i.e., the acceptable number of unnecessary biopsies to perform to catch an event (disease progression). The left and right columns summarize the results where $T_{\text{gap}}=24$ and $T_{\text{gap}}=48$, respectively.}
	\label{fig:sim2}
\end{figure}

Figures~\ref{fig:sim1} and~\ref{fig:sim2} summarize the results for Scenarios~(1) and~(2), respectively. In both scenarios, compared to the OSF-R approach, the proposed OSF-I method achieves higher values for almost all choices of $r$. Compared with the settings where $T_{\text{gap}}=24$, in the settings where $T_{\text{gap}}=48$, the advantage of the proposed method is larger; this implies that the bias induced by treating the interval-censored data as right-censored increases with the increase of $T_{\text{gap}}$. 

\section{Application}
\label{sec:real}

In this section, we apply the proposed method to develop and evaluate a clinical decision rule for the tailored management of prostate cancer patients using data from PASS. We included 844 patients diagnosed since 2003 and enrolled in PASS before 2017, with Gleason grade group $(GG)$ 1 on diagnostic biopsy and GG1 or no tumor on confirmatory biopsy. The disease progression was defined as a reclassification, any increase in GG to 2 or more, detected through a surveillance biopsy. The predictors included the most recent PSA values, the most recent BMI status (normal, overweight, or obesity), the logarithm of the most recent prostate size, the PSA value at diagnosis, the most recent maximum core ratio, the logarithm of time since the confirmatory biopsy, and the counts of negative biopsies $(0,1,\geq 2)$.

We aim to derive an AS rule based on updated information to decide whether a patient should be examined with a biopsy within one year ($\tau$ = 1 year). The time points of the decisions were chosen to be $s$ years after the confirmatory biopsy, where $s=1,2,3,4$. To compare different methods, we conduct two analyses. In the first analysis, we use a repeated sample-splitting strategy and consider only data from PASS cohort. In this analysis, we first split the entire PASS cohort into a training dataset and a testing dataset with equal sample sizes. Then we implement each method on the training dataset and calculate the TPR, TNR, and the weighted benefits value on the testing dataset. We report the mean and standard deviation of these summary measures over $100$ repeats. In the second analysis, we implement each method on the entire PASS cohort and use an external cohort from the University of California San Francisco (UCSF) to evaluate the TPR, TNR, and the weighted benefits value of the decision rule derived on the PASS cohort. To construct confidence intervals, we use bootstrap with bootstrap number $B=1000$. Table~\ref{tab:cohort} summarizes the patient characteristics of the PASS and UCSF cohort. We observe that compared with the PASS cohort, the UCSF cohort is younger, more diverse, and has a higher event rate (grade reclassification).

\begin{figure}
	\centering
	\includegraphics[width=0.5\linewidth]{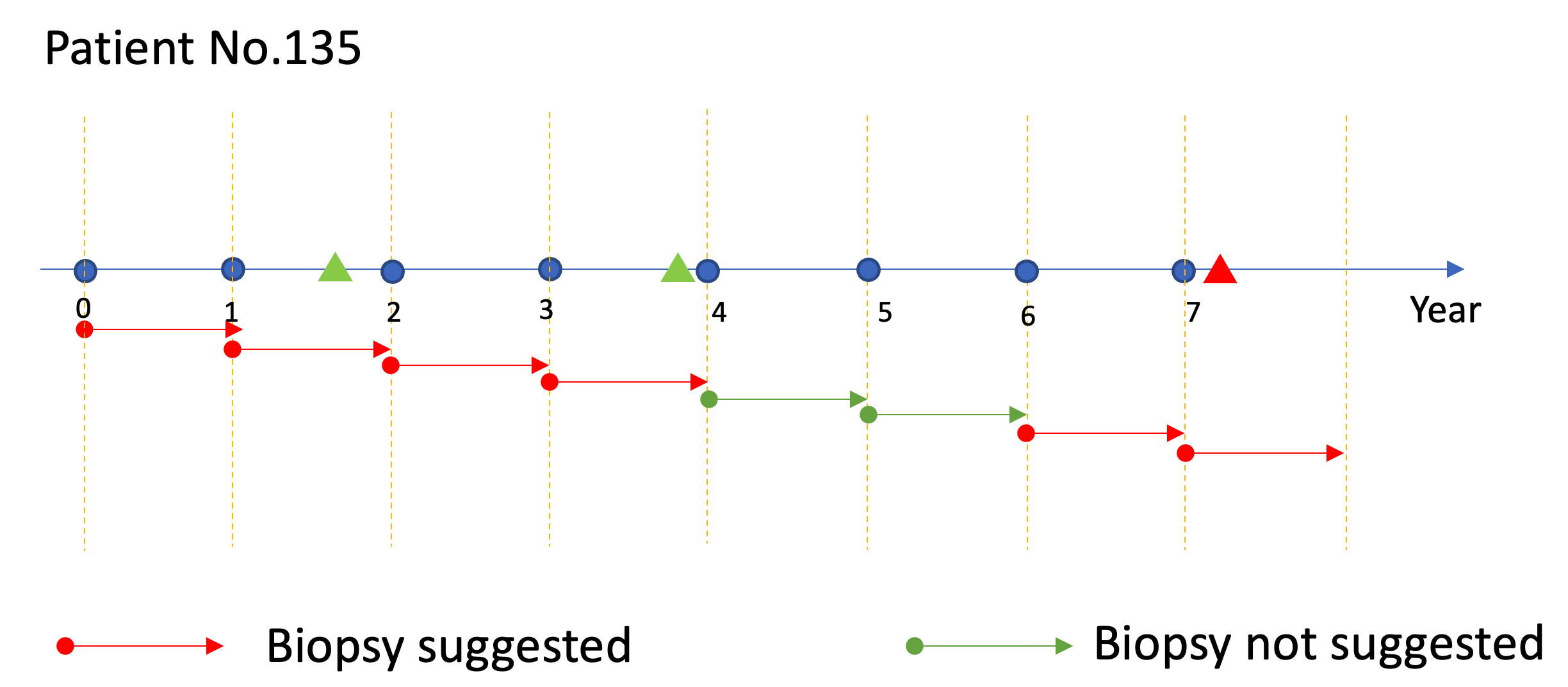}
	\caption{Visualization of the estimated AS strategy for Patient No.135. For this patient, there are three biopsies recorded in the PASS data (green triangular represents the biopsy that detects no disease progression; red triangular represents the biopsy detecting disease progression), and there is a disease progression detected at Year 7.1. The blue points represent the time updating covariate information; the arrows below the time axis represent our estimated AS decisions at Years 0 -~7 (apply the stabilized AS decision rule derived from Year 1-~4 in Years 0 -~7). Our estimated strategy suggests that the patient should skip biopsies at Years 4 and 5 due to the increased prostate size, and resume the biopsies at Years 6 and 7 due to abnormally increased PSAs.}
	\label{fig:visualization}
\end{figure}

In both analyses, we set up a sequence of cost-benefit ratios ranging from $4$ to $12$. We consider a wide range of cost-benefit ratios reflecting varied emphasis on the TPR (increase from about $10\%$ to higher than $90\%$ when using the PASS cohort). We used the repeated sample-splitting strategy to compare different methods. Table~\ref{tab:compare_real_data} reports the results for both analyses. When the cost-benefit ratio ($r$) is 4, the proposed OSF-I method achieves comparable or slightly higher value when compared with the OSF-R method proposed under right-censoring; when the cost-benefit ratio ($r$) is larger than 6, the proposed OSF-I method achieves significantly higher value than the OSF-R method. When we use the UCSF data to validate the tailored AS rules derived by different methods, although the confidence intervals, in this case, are large, we can still observe that the proposed OSF-I method tends to achieve higher values than the OSF-R method. We further visualizes our estimated strategy to make biopsy decisions (see Figure~\ref{fig:visualization}).


\section{Discussion}
\label{sec:discuss}
In our research, we propose a weighted classification approach to determine the optimal Active Surveillance (AS) strategy in scenarios involving interval-censored events and immediate dropouts following a positive biopsy. Our approach differs from existing methods used for right-censored or panel status data without immediate dropouts. We utilize adjacent negative-positive pairs and employ two-dimensional kernel regressions for estimating True Positive Rates (TPRs) and True Negative Rates (TNRs) to accommodate the complications of interval-censored events and immediate dropouts. 

Our work opens several promising avenues for future investigation. Firstly, while our current AS strategy assumes up-to-date patient information, real-world scenarios often involve missing data due to patient non-adherence to study protocols. To address this, employing techniques like the 'last value carry-forward' method to impute missing data from previous time points could be considered. However, integrating such imputed data may impact the optimality of the derived AS strategy. Exploring methods to incorporate delayed or outdated information into AS strategy formulation would be a valuable pursuit. Secondly, our study relies on biopsies to identify disease progression. However, biopsies may have imperfect sensitivity or specificity in detecting progression. Third, we make a random censoring assumption, which may not hold for patients opt for treatment before a positive biopsy. Deriving AS strategies that account for these imperfections in disease detection is an intriguing area for future research. Continuing exploration in these directions could significantly enhance the practical applicability and robustness of AS strategies in clinical settings.

\begin{sidewaystable}
\centering
\caption{Patient characteristics of the PASS and UCSF cohort}
\label{tab:cohort}
\begin{tabular}{lcc|cc}
\hline
Variable & \multicolumn{2}{c|}{PASS (844 Patients)} & \multicolumn{2}{c}{UCSF (533 Patients)} \\
\hline
Age, No. (\%), year & & & & \\
\quad $<$60 & 290 (34) & & 222 (42) & \\
\quad 60-70 & 474 (56) & & 271 (51) & \\
\quad $>$70 & 80 (10) & & 40 (7) & \\
BMI, median (IQR) & 27 (25 to 30) & & 27 (25 to 29) & \\
Race/ethnicity, No. (\%) & & & & \\
\quad White & 769 (91) & & 422 (79) & \\
\quad Black & 42 (5) & & 12 (2) & \\
\quad Other & 33 (4) & & 99 (19) & \\
Diagnostic percent positive cores, median (IQR),\% & 8.3 (8.3 to 16.7) & & 11 (7 to 19) & \\
No. missing percent positive cores at diagnosis & 16 & & 7 & \\
Diagnostic PSA, median (IQR), ng/mL & 4.7 (3.5 to 6.4) & & 5.4 (4.2 to 7.3) & \\
No. PSA measurements, median (IQR) & 12 (7 to 19) & & 7 (4 to 13) & \\
Most recent prostate size at confirmatory bx, median (IQR), mL & 42 (30 to 58) & & 39 (30 to 54) & \\
Grade reclassification, No. (\%) & 182 (22) & & 154 (29) & \\
Follow-up since confirmatory bx, censored patients, median (IQR), y & 3.2 (1.7 to 5.0) & & 2.5 (1.3 to 4.3) & \\
\hline
\end{tabular}
\end{sidewaystable}

\begin{sidewaystable}
\centering
\caption{Comparisons using the PASS and UCSF data}
\label{tab:compare_real_data}
\begin{tabular}{c|ccccccccc}
\hline
\multicolumn{7}{c}{PASS Only}\\\hline
$r$ & & 4 & 6 & 8 & 10 & 12\\
\hline
\multirow{3}{*}{OSF-I}& TPR & 0.120 (0.100,0.140) & 0.489 (0.456,0.522) & 0.770 (0.744,0.796) & 0.893 (0.875,0.910) & 0.946 (0.934,0.958)\\
& TNR & 0.951 (0.942,0.960) & 0.730 (0.707,0.754) & 0.442 (0.414,0.471) & 0.244 (0.218,0.270) & 0.140 (0.122,0.158)\\
 & Value & {1.196 (1.184,1.208)} & {1.038 (1.020,1.055)} & 1.017 (1.003, 1.031) & 1.002 (0.992, 1.011) & 0.998 (0.991, 1.004)\\\hline
\multirow{3}{*}{OSF-R}& TPR & 0.049 (0.037,0.061) & 0.191 (0.170,0.211) & 0.344 (0.318,0.369) & 0.483 (0.460,0.506) & 0.606 (0.584,0.627)\\
& TNR & 0.984 (0.980,0.987) & 0.934 (0.927,0.941) & 0.861 (0.850,0.873) & 0.772 (0.759,0.785)& 0.676 (0.663,0.690)\\
 & Value & 1.162 (1.153,1.171) & 0.896 (0.880,0.913) & 0.832 (0.811,0.852) & 0.833 (0.815,0.852)& 0.862 (0.844,0.879)\\\hline
\multicolumn{7}{c}{PASS Train + UCSF Test}\\\hline
$r$ & & 4 & 6 & 8 & 10 & 12\\
\hline
\multirow{3}{*}{OSF-I}& TPR & 0.180 (0.128,0.303) & 0.633 (0.528,0.735) & 0.803 (0.700,0.881) & 0.858 (0.794,0.940) & 0.920 (0.858,0.981)\\
& TNR & 0.901 (0.812,0.900) & 0.496 (0.429,0.567) & 0.288 (0.233,0.357) & 0.198 (0.139,0.247) & 0.119 (0.054,0.128)\\
 & Value & 0.805 (0.666,1.072) & 0.861 (0.744,1.008) & 0.901 (0.803,0.994) & 0.910 (0.848,0.995) & 0.946 (0.882,1.003)\\\hline
\multirow{3}{*}{OSF-R}& TPR & 0.055 (0.015,0.112) & 0.279 (0.193,0.376) & 0.497 (0.348,0.552) & 0.619 (0.510,0.718) & 0.697 (0.595,0.791)\\
& TNR & 0.971 (0.950,0.987) & 0.816 (0.767,0.863) & 0.619 (0.615,0.740) & 0.502 (0.437,0.568) & 0.412 (0.350,0.480)\\
 & Value & 0.728 (0.582,1.013) & 0.660 (0.545,0.843) & 0.718 (0.588,0.837) & 0.762 (0.657,0.881) & 0.795 (0.699,0.899)\\\hline
\end{tabular}
\end{sidewaystable}

	\newpage
	
	\begin{center}
		{\large\bf SUPPLEMENTARY MATERIALS}
	\end{center}
	
	\begin{description}
		
		\item[Supplemental Materials:] An R package implementing the proposed method can be downloaded at \url{https://github.com/muxuanliang/CancerSurv.git}. 		
	\end{description}

\singlespacing
\bibliographystyle{apalike}
\bibliography{ref.bib}

\end{document}